# Highly directional and coherent emission from dark excitons enabled by bound states in the continuum

Xuezhi Ma[1], Kaushik Kudtarkar[1], Yixin Chen[2,3], Preston Cunha[1], Yuan Ma[1,4], Kenji Watanabe[5], Takashi Taniguchi[6], Xiaofeng Qian[3], M. Cynthia Hipwell[1], Zi Jing Wong[2,3], and Shoufeng Lan[1,3*]

1 Department of Mechanical Engineering, Texas A&M University, College Station, TX 77840, USA.

2 Department of Aerospace Engineering, Texas A&M University, College Station, TX 77840, USA.

3 Department of Materials Science and Engineering, Texas A&M University, College Station, TX 77840, USA.

4 Department of Mechanical Engineering, The Hong Kong Polytechnic University, Hong Kong, China.

5 Research Center for Functional Materials, National Institute for Materials Science, Tsukuba, Japan.

6 International Center for Materials Nanoarchitectonics, National Institute for Materials Science, Tsukuba, Japan.

Author e-mail address: shoufeng@tamu.edu

**Abstract**

A double-edged sword in two-dimensional material science and technology is an optically forbidden dark exciton. On the one hand, it is fascinating for condensed matter physics, quantum information processing, and optoelectronics due to its long lifetime. On the other hand, it is notorious for being optically inaccessible from both excitation and detection standpoints. Here, we provide an efficient and low-loss solution to the dilemma by reintroducing photonic bound states in the continuum (BICs) to manipulate dark excitons in the momentum space. In a monolayer tungsten diselenide under normal incidence, we observed a giant enhancement with an enhancement factor of ~3,100 for dark excitons enabled by transverse magnetic BICs with intrinsic out-of-plane electric fields. By further employing widely tunable Friedrich-Wintgen BICs, we demonstrated highly directional emission from the dark excitons with a divergence angle of merely 7 degrees. We found that the directional emission is coherent at room temperature, unambiguously shown in polarization analyses and interference measurements. Therefore, the BICs reintroduced as a momentum-space photonic environment could be an intriguing platform to reshape and redefine light-matter interactions in nearby quantum materials, such as low-dimensional materials, otherwise challenging or even impossible to achieve.

Dark excitons ($X_D$) in semiconductors have a long lifetime due to their decoupling from radiative channels and spin-flip processes[1-5], making them perfect to serve as the quantum bits (qubits) for the ongoing development of quantum computing[6]. Dark excitons in two-dimensional (2D) semiconductors, such as transition metal dichalcogenide (TMD) monolayers, have also attracted broad research interests thanks to the atomically thin structure of 2D materials that made them suitable for compact planar devices[7]. However, since the excitonic transition of dark excitons does not satisfy the selection rules and possesses zero in-plane dipole moments[3,8,9], achieving dark exciton optical brightening with conventional far-field optical techniques has been a challenging task[10]. Additionally, collecting $X_D$ emission requires a large-numerical-aperture (NA) objective lens due to their out-of-plane dipole-like radiation pattern, which made it challenging to achieve information read-out using conventional far-field optical techniques. To demonstrate a robust dark exciton qubit read-out system, two tasks need to be satisfied, namely dark exciton optical brightening, and efficient collection of the $X_D$ emission.

For dark exciton brightening, firstly, the current techniques to excite the $X_D$ emission usually requires an ultra-strong in-plane magnetic field ($> 14T$)[2] or out-of-plane polarized surface plasmon polaritons (SPPs)[3]. The former brightening method uses an ultra-strong in-plane magnetic field to tilt the effective internal magnetic field in the conduction band (CB) to gain an in-plane component from the original out-of-plane electron spin. The latter method uses the SPP structure to convert the in-plane polarized incident light into an out-of-plane SPP mode to efficiently couple with the out-of-plane transition dipole moment of dark excitons. In addition, out-of-plane transition dipole moments of dark excitons in $WSe_2$ monolayer encapsulated between thin h-BN flakes can couple with out-of-plane polarized incident light provided by a 90-degree rotated objective lens, i.e., the dark excitons can be excited by the horizontally oriented objective lens[10]. These methods, however, are intrinsically restricted to cryogenic temperature conditions (T < 30K). Alternatively, a tip-enhanced photoluminescence (TEPL) system manifested itself as capable of brightening dark excitons at room temperature[4]. The oblique incident laser can be coupled in, and its out-of-plane polarized component can be selectively enhanced by the scanning probe microscope (STM) driven gap mode between the gold tip and gold substrate, yielding a greater coupling efficiency with the out-of-plane transition dipole moment of dark excitons. However, its complicated setup limits its application, especially when bringing the dark exciton read-out system to an on-chip systems[11].

Secondly, efficient $X_D$ emission collection is another bottleneck for the qubits read-out system. A grating coupler[3], waveguide-based method[9], or oblique collecting setups[4] are usually used to boost the collection efficiency of $X_D$ emissions. However, the complex setup, relatively low collection efficiency, or the absence of the enhancement of $X_D$ emission limit their applications. On the other hand, optical resonators such as photonic crystal cavities[12], whispering gallery mode (WGM) resonators[13], antenna array Mie resonators[14], and metamaterials[15] are used to enhance and directionally emit the electroluminescence (EL) or photoluminescence (PL) signals from bright excitons. To combine those advantages, designing a structure that can achieve the directional $X_D$ emission with suitable deflection angles can break the bottleneck for the $X_D$ read-out system. In this way, $X_D$ emission can be easily detected by either a conventional microscope or a single detector. Practically, recently emerged optical bound states in the continuum (BICs) with infinite quality-factors (Q-factors) or quasi-BICs with finite Q-factors stand out from optical resonators for this mission, thanks to their high Q-factor and compatibility with planar optical platforms. These BICs can be supported in various photonic systems such as photonic crystal slabs[16], plasmonic structures[17], metasurfaces[18], and fiber Bragg gratings[19]. BICs can be assigned to three categories: (i) symmetry-protected BICs at Gamma-points (Γ-points, the center of the Brillouin zone), (ii) off-Γ accidental BICs, and (iii) Friedrich-Wintgen BICs[20,21]. In 1985, Friedrich and Wintgen suggested that BIC can occur due to the interference of resonances belonging to different channels that cause an avoided crossing[22]. The avoided crossing was then extended to acoustic[23], quantum[24] and optical systems[25]. Compared with accidental BICs, which rely on carefully tuning the geometric parameters, Friedrich-Wintgen BICs are stable and efficient, making them perfect candidates for $X_D$ directional emission[26,27].

In the present work, we designed a suspended photonic crystal (PhC) slab made of lossless silicon nitride ($Si_3N_4$) which simultaneously supports on-Γ symmetry-protected BICs and off-Γ Friedrich-Wintgen BICs (**Fig. 1a**). In this design, a transverse magnetic (TM) like BIC at the Γ-point can efficiently convert the in-plane polarized normally-incident pump laser into out-of-plane polarized near-field energy to gain significant efficiency to couple with the out-of-plane transition dipole moment of dark excitons[11,16,28]. In this way, the forbidden electrons transitioning from the valence band (VB) to the conduction band (CB) with opposite spin direction can be allowed, resulting in dark exciton brightening (**Fig. 1b**, the $X_D$ transitions show this process). Meanwhile, the off-Γ Friedrich-Wintgen BIC can selectively couple with the out-of-plane transition dipole moment of dark excitons and directionally emit the $X_D$-PL signals[29,30]. In the experiments, we transferred exfoliated monolayers of $WSe_2$ that were covered by a few-layer thick hexagonal boron nitride (h-BN) flake onto the suspended $Si_3N_4$ PhC slab device (**Fig. S2,** the optical microscope picture of the device and **Fig. 1c**, the scanning electron microscope image of the PhC slab, and the transfer details see Method). The sketch of the band structure of the device, as shown in **Fig. 1d**, was extracted from the numerically simulated angle-resolved reflection spectroscopy mapping to highlight the BICs modes supported by the designed PhC slab. It clearly shows three types of BICs: ① and ④ are on-$\Gamma$ symmetry-protected BICs, ② is an off-$\Gamma$ accidental BIC, and ③ is an off-$\Gamma$ Friedrich-Wintgen BIC due to the destructive interference of resonances belonging to different TM-like bands (red and blue). In our device, the PhC slab fulfills two important roles simultaneously: it converts polarization and enhances the incident light, and it selectively enhances and directionally emits the $X_D$-PL signal. Only with these two roles achieved at the same time can the WSe$_2$ monolayer integrate with photonic chips for robust dark exciton qubits read-out system. With the PhC slab, the double BICs mediated far-field-to-near-field-to-far-field mode transformation, giving rise to a ~ 3,100-fold $X_D$-PL enhancement demonstrated in the WSe$_2$ monolayer with a Purcell factor of 26.7. Furthermore, the Friedrich-Wintgen BIC can be tuned to emit the dark excitons signal in various angles from 31.4° to 59.5° by tuning the device parameters, demonstrating an angle-tunable dark exciton read-out system. Finally, we also demonstrate the $X_D$ directional emissions are coherent at room temperature.

**Dark exciton brightening using on-$\Gamma$ symmetry protected BIC at room temperature.** To demonstrate the mechanism of dark exciton brightening by on-$\Gamma$ symmetry protected BIC, we firstly designed a suspended Si$_3$N$_4$ PhC slab (n = 2.23, thickness 265 nm) with a square array of cylindrical holes (periodicity 450 nm, hole diameter 140 nm). The slab material, Si$_3$N$_4$, is deposited using low-pressure chemical vapor deposition (LPCVD) on top of an <100> oriented silicon wafer, providing low absorption and the ability to guide the electromagnetic energy inside. The optimized fabrication protocol (see Method) was applied to yield a PhC slab with minimum surface roughness to minimize scattering and improve the quality of the interface between the WSe$_2$ monolayer and the PhC slab. The device was 100 by 100 periods (45x45um$^2$), which was sufficiently large to guarantee the C$_4$ symmetry and periodic boundary condition of the lattice in the central region of the device. **Fig. S2a** shows the transferred WSe$_2$ monolayer with a few-layer h-BN flake covering the suspended PhC slab. The h-BN flake can protect the WSe$_2$ monolayer from oxidation, especially under the strong pump laser (several mW or higher) in the air at room temperature[31]. It is worth mentioning that the h-BN flake can introduce a red-shift in spectra to the PhC slab[32]. To compensate for this drift, we reduced the thickness of the PhC device by several nanometers in our experiment to minimize any spectral differences between the simulation and the real device. This compensation strategy is effective, thanks to the small difference in refraction index between our Si$_3$N$_4$ slab (2.23) and the exfoliated h-BN flake (~ 2.2, near 700 nm in wavelength)[33].

To better understand the optical properties of the PhC device, we performed numerical simulations and characterizations of the device. **Fig. 2a** shows the dispersion curves of the eight lowest energy bands (optical band structures, by the MPB band solver[34]) along the Γ-X line [$k(\Gamma) = (0,0)*(2\pi/a), k(\Gamma) = (0.5,0)*(2\pi/a), k = (k_x + k_y)$ and $k_x = (\omega/c)\sin(\theta)$]. The four blue bands are TM-like, whereas the four red bands are transverse electric (TE) like. Three TM-like bands and one TE-like band can be excited by the p-polarized incident light as shown in the mapping of the reflection spectra as a function of incident angle in **Fig. 2b.** The comparison between the numerical simulation (left) and the angle-resolved spectrometric measurement (right) shows minimum differences, demonstrating the quality of the fabricated device. Two symmetry-protected TM-like BIC modes at the Γ-point at 694 nm and 755 nm are clearly visible in **Fig. 2b**. We designed a BIC at 694 nm as the incident light polarization converter because the BIC can convert the in-plane polarized incident light into the out-of-plane polarized component with a large enhancement ratio, as shown in **Fig. 2c** (the out-of-plane electric field component, $E_z$ distribution on the X-Y plane (upper) and the X-Z plane (lower) in the WSe$_2$ monolayer). A numerical simulation using COMSOL Multiphysics shows that $E_z$ can be created and enhanced with an enhancement ratio reaching ~ $3\times10^4$-fold at the maximum point of resonance (BIC) when compared with the original incident electrical field. In the real experiment, the tightly focused mono-color incident laser spot still has a relatively large divergent incident angle and a wide width of wavelength in the spectra. As a result, only a small part of the energy of the incident laser can couple with the BIC and be converted and enhanced, even if the 5X beam expander (Thorlabs, BE05-10-A was used to compress the beam size, more details see Supplementary section 8, **Fig. S8a**) and the laser line filter (Thorlabs, FL694.3-10 was placed with a small twisted angle, more details see Supplementary section 8, **Fig. S8b** and **c**) were applied to compress the laser energy into the BIC mode. **Fig. 2d** shows the reflection spectrum with an oblique incident angle of 3° (indicated by the black dashed line) and the maximum local $E_z$ enhancement ratio over the incident light electric field on the top surface of the PhC slab, where the WSe$_2$ monolayer is seated. However, it is clearly seen that only the laser energy that is close to the resonance peak benefits from the large enhancement factor. To quantitatively estimate the $E_z$ conversion and enhancement efficiency, we calculated the average $(E_z/E_0)^2$ as the enhancement factor (more details see Supplementary section 8) and found that the average $(E_z/E_0)^2$ is 216 when the full-width-half-maximum (FWHM) of the wavelength is 5 nm and the FWHM of the incident laser divergence angle is 4° (+/- 2°), which provides sufficiently large enhancement of the out-of-plane component of the near field energy for dark exciton brightening.

The PL spectra of the WSe$_2$ monolayer device excited at a wavelength of 694 nm (blue, on-BIC) and 647 nm (red, off-BIC), is shown in **Fig. 2e**. We used a 40X large numerical aperture objective lens (40X Nikon S Flour high NA objective, NA = 0.9) to excite the incident laser and collect the PL signal. A strong X$_D$ emission peak was observed (blue spectrum) when the incident laser matched the BIC mode, whereas the X$_D$ emission channel was closed when the incident laser was off-resonance (off-BIC, red spectrum). From the PL spectra for X$_D$ and bright excitons (X$_0$), we obtained ~ 52 meV of intravalley energy splitting between the X$_D$ and X$_0$. This result is in good agreement with other X$_D$ emission observations, namely 47 meV obtained by in-plane magnetic field[2], 42 meV by SPP coupling[3], and 46 meV by TEPL setup[4]. To eliminate the possibility that emission stemmed from bi-excitons, which have similar emission energy compared to dark excitons in WSe$_2$[35], we measured the PL signal intensity as a function of excitation power. The intensity of the PL emission and the excitation power are related, i.e., $I_{PL} \propto I_{excited}^{\alpha}$, with α = 2 for bi-excitons and α = 1 for bright or dark excitons, respectively. In a realistic experiment, the factor α would be expected in the range of 1.2-1.9 for bi-excitons, which have been extensively studied[35,36]. In our experiment, the logarithmic plot of the dark excitons PL peak with respect to the excitation power, as shown in **Fig. 2f**, shows the fitted exponent of α = 0.9, indicating the emission stemmed from dark excitons.

**Directional and coherent emission of dark excitons by Friedrich-Wintgen BIC.** Because of their out-of-plane dipole radiation nature, X$_D$ emit primarily towards the in-plane direction, making the PL signal difficult to collect with conventional microscopic objective lenses[10]. **Fig. S10** shows the collection efficiency of PL emission radiated by dark excitons and bright excitons, respectively. It is clear to see barely any of the X$_D$ emission can be collected by an objective lens with a sufficiently large numerical aperture (NA = 0.9). To gain higher dark exciton read-out efficiency, we demonstrated a directional and polarized emission channel with tunable angles for dark excitons using the Friedrich-Wintgen BICs. This directional emission channel can selectively couple with the out-of-plane dipole moment of dark excitons and enhance their directional emission. The enhanced directional emission has a small full width at half-maximum (FWHM) of ~ 7° divergence angle in air and a ~ 3,100-fold total enhancement factor with linear polarization. By tuning the geometric parameters of the PhC slab, the Friedrich-Wintgen BIC can be tuned to provide different deflection angles for dark exciton emission, making dark exciton read-out possible by a single detector placed in the proper position.

In this experiment, the Friedrich-Wintgen BIC serves as a stable resonator with a high Q-factor and provides an efficient radiation channel for $X_D$ directional emission[30,37]. We designed a suspended $Si_3N_4$ PhC slab (n = 2.23, thickness 233 nm) with a square array of cylindrical holes (periodicity 510 nm, hole diameter 102 nm) that supports a TM-like symmetry protected BIC at the Γ-point of 596 nm for dark exciton brightening (**Fig. S5**) and another TM-like Friedrich-Wintgen BIC at $k_x$ = 0.74 (47.85°) and 770 nm for the $X_D$ emission (**Fig. 3a** and **b**). **Fig. 3a** and **Fig. S4** show the simulated and measured mapping of the reflection spectra as a function of incident angle (optical band structure), respectively. It is clear to observe the avoided crossing in the white dashed line labeled region where the Friedrich-Wintgen BIC occurs in **Fig. 3a**. **Fig. 3b** shows the zoomed-in optical band structure in the Friedrich-Wintgen BIC region, indicating that the Friedrich-Wintgen BIC occurs due to the destructive interference between two TM-like bands (Mode analyses see **Fig. S7**). In the reflection spectra (**Fig. 3c**), each mode can be described by an asymmetrical Fano line shape. Vanishing of the Fano lineshape occurred at the lower branch at 47.85°, indicating that the BIC emerged with an infinite Q-factor via the destructive interference of the two resonances. We used a numerical simulation (COMSOL Multiphysics) eigenfrequency model to extract the Q-factors along each of the two bands as shown in **Fig. 3d**. The Q-factor of the upper branch remains relatively low (~ $10^3$) whereas the Q-factor of the lower branch has an infinite peak at around 47.85°.

In addition, the Friedrich-Wintgen BICs are momentum tunable, and thus we can control the $X_D$ emission towards different deflection angles. We designed a series of the Friedrich-Wintgen BICs at the wavelength of 770 nm in **Fig. 3e**. This momentum tunability of the dark exciton's directional emission channel, i.e., the Friedrich-Wintgen BIC, can also benefit dark exciton information read-out using single detectors. The Friedrich-Wintgen BICs stem from the full destructive interference of two TM bands thus are stable and easier to design in comparison with the off-Γ accidental BIC. We found that the avoided crossing of the two TM bands, exactly where the Friedrich-Wintgen BIC occurs, can be tuned by varying the PhC slab's thickness. It is clear to see, as shown in **Fig. 3e**, the avoided crossing moved towards a higher angle in the momentum space as the slab thickness increased. To match the dark exciton energy (770 nm in wavelength), we changed the periodicity and as well as the radius of holes for each PhC design. The continuum tunable angle ranges from 31.4° to 59.5° as we demonstrated for some discrete cases. The white dashed line traces the avoided crossing to show the tuning trend of the Friedrich-Wintgen BICs. We further plotted the deflection angle by Friedrich-Wintgen BICs as the function of the PhC slab thickness. The relationship was quasi-linear, implying that we can extend the tunable deflection angle to a broader range.

The directional and polarized emission of dark excitons through the Friedrich-Wintgen BIC has been analyzed in **Fig. 4**. The few-layer h-BN flake and monolayer WSe$_2$ heterostructure were sequentially transferred onto the PhC slab that supported the Friedrich-Wintgen BIC and pumped by a wavelength-tunable femtosecond-ultrafast-laser with the power of 50 μW with p-polarization. **Fig. 4a** shows the PL emission momentum distribution mapped with $k_x$ and $k_y$ axis. Four shining emission spots, located at the Γ-X line with $k_x/k$ or $k_y/k = 0.74$ and with C$_4$ symmetry, are clearly visible. The PL emission momentum distribution was collected by a sensitive silicon camera (Princeton Instruments PIXIS 400) that was placed in the K-plane (see Method). It is worth noting that both X$_0$ and X$_D$ emissions were collected by the camera at the same time. To elucidate the PL emission momentum distribution in further depth, the polarization-resolved images are presented in **Fig. 4b**. The four shining spots were radially polarized and exhibit similar intensities. Three pieces of evidence are present to explain why the origin of these four shining spots comes from the dark excitons. 1) **Fig. S6** shows the PL emission polarization analysis of the bright or dark excitons, and it shows that only dark excitons' out-of-plane dipole radiation nature can provide the emission with C$_4$ symmetry (radial polarization); 2) The four shinning spots have similar emission intensity. If those spots came from bright excitons with p-polarized (along with the y-direction) in-plane momentum, the spots along the x-direction should be brighter than those along the y-direction. This argument has already considered the depolarization effect of bright exciton at room temperature (more details see Supplementary section 12); 3) Only dark excitons with out-of-plane dipole momentum have high coupling efficiency with the Friedrich-Wintgen BIC that comes from the interference between two TM-like bands. (For the E-field analysis of the BIC mode see **Fig. S7**). Furthermore, the polarization for each directional emission spot is linear and the extinction ratio is relatively high (shining spot can be fully blocked under cross-polarizer), proving its coherence at room temperature[38].

The angle-resolved PL emission spectra mapping extracted from the PL emission momentum distribution with p-polarization (y-direction) by the diffraction grating of the spectrometer is shown in **Fig. 4c**. The enhanced X$_D$ emission located at $k_y/k = +/-0.74$ at 772 nm with a small divergence angle is clearly visible (~ 7°, **Fig. 4d**). We further extracted the spectra from the angle-solved PL emission as shown in **Fig. 4e**, the blue line of which shows the X$_D$ emission spectrum extracted from $k_y/k = 0.74$ and the red line shows the bright excitons emission spectrum from $k_y/k = 0.39$. This is clear evidence that the PhC slab can separate the emission from bright and dark excitons and selectively enhance the X$_D$ emission at a specified angle with a small divergence angle.

To further understand the coupling mechanism between the out-of-plane dipole and the Friedrich-Wintgen BIC supported by the PhC slab, a Finite-difference time-domain (FDTD) simulation was performed using commercial software (Lumerical FDTD Solutions, ANSYS Inc.) to show how the PhC slab selectively couples with the out-of-plane dipole and enhances their emission. A series of out-of-plane dipoles with excited wavelengths near 770 nm (on-resonance of the Friedrich-Wintgen BIC) were randomly placed on the top surface of the PhC slab (100 by 100 periods with scattering boundary conditions). **Fig. 4f** shows the simulated out-of-plane dipole emission momentum distribution, which perfectly matched with the $X_D$ emission part of the PL emission momentum distribution (**Fig. 4a**). The broadening of the emission features in the azimuthal direction of the $X_D$ emission can be attributed to the thermal expansion at room temperature.

We also quantitatively estimate the enhancement factor (EF) of the $X_D$ emission by the Friedrich-Wintgen BIC performed by the FDTD simulation (more details see Method). The EF equation is $EF = \left|\frac{E_z}{E_0}\right|^2 \times \gamma_{PF}$, where $\gamma_{PF}$ is the Purcell factor of the Friedrich-Wintgen BIC. We used the FDTD method to calculate the Purcell factor the highest of which is 26.7 at a wavelength of 769 nm (more details see Supplementary section 9). Because of both the monolayer $WSe_2$ flake size and the tightly focused incident light spot is larger than one period of the PhC lattice (510 nm), we used the average $|E_z/E_0|^2$ enhancement ratio instead of the near-field E-field enhancement ratio at any local point to estimate the EF. The average $|E_z/E_0|^2$ enhancement ratio by the on-Γ BIC at 596 nm is 116 (more calculation details see Supplementary section 8). The EF is estimated to be as high as ~ 3,100 at a wavelength of 769 nm.

**Coherence demonstration of directional emission from dark excitons.** Due to the long lifetime of the dark exciton and the high Purcell factor of the Friedrich-Wintgen BIC cavity, the directional emission of the dark excitons in monolayer $WSe_2$ maintains their coherent nature[39]. To demonstrate this, we used a cylindrical lens to focus the emission pattern to observe its interference pattern[40,41](more details see Supplementary section 12). Before the coherence demonstration, we used a laser spot of Gaussian shape, reflected by a silicon wafer to observe the bright exciton emission pattern in K-space from monolayer $WSe_2$ on a $SiO_2$/Si substrate to demonstrate this method's reliability. **Fig. 5a, c and e** respectively show the K-space images (left) and the intensity profiles (right) of the $X_D$ directional emission, laser spot, and bright exciton emission, where **Fig. 5b, d, and f** show their corresponding cylindrically focused patterns (left) and intensity profiles (right). Because the bright exciton emission would lose its valley coherence at room temperature and shows nearly random polarization, the intensity profile (labeled by dashed line) of the K-space pattern (**Fig. 5e**, right) and the cylindrically focused pattern (**Fig. 5f**, right) should be Gaussian-like[4]. On the other hand, the laser spot has naturally good coherence and as a result, its intensity profile of the cylindrically focused pattern (**Fig. 5c**, right) should have a narrower FWHM because of the constructive interference of the common phase. Finally, we cylindrically focused the dark exciton directional emission pattern into one dimension, meaning the left and right emission spots are overlap. Because they (the left and right spots) have the same polarization and a phase difference of $\pi$ ($C_4$ symmetry of this system) as we discussed in the manuscript, the destructive interference pattern in the middle is expected (**Fig. 5b**).

**Conclusion**

We designed a suspended $Si_3N_4$ PhC slab that simultaneously supports a TM-like on-$\Gamma$ symmetry-protected BIC and a tunable off-$\Gamma$ Friedrich-Wintgen BIC. The supported on-$\Gamma$ BIC can efficiently convert the in-plane polarized normal incident pump laser into out-of-plane polarized near-field energy to gain large efficiency as high as 116 to couple with the out-of-plane transition dipole moment, thus brightening the optically forbidden dark excitons in monolayer $WSe_2$ at room temperature. We also demonstrated that the pump laser could regulate the $X_D$ emission by switching frequencies on- or off- the BIC mode. Further, the off-$\Gamma$ Friedrich-Wintgen BIC can selectively couple with the out-of-plane transition dipole moment of dark excitons and provide an efficient directional and polarized emission channel for the dark excitons with minimized divergence angle (FWHM of ~ 7° in the air) and ~ 3,100-fold enhancement factor. By tunning the thickness of the PhC slab, the Friedrich-Wintgen BIC can be tuned to emit the dark excitons signal at various angles from 31.4° to 59.5°. Furthermore, we demonstrated the coherence of the $X_D$ directional emission at room temperature by the constructive interference of the directional emissions that towards opposite directions. The dark exciton brightening and directional and coherent emission by a planar device demonstrate a robust dark exciton information read-out system, paving the way for on-chip computing and communications.

**Method**

**Sample fabrication**

A 300 nm thick layer of $Si_3N_4$ was deposited on Si substrates using low-pressure chemical vapor deposition (LPCVD). The periodic PhC pattern with periodicity of 465 nm (or 510 nm) and hole radius of 140 nm (or 102 nm) was written on ZEP-520A photoresist using the TESCAN MIRA3 E-beam lithography system. Then the RIE-ICP cyclic dry etch strategy was applied to transfer the pattern from the photoresist to the $Si_3N_4$ layer. After removing the ZEP-520 photoresist residual with hot acetone (90 °C) and an oxygen plasma cleaning procedure, the silicon (<100> oriented) beneath the PhC pattern was undercut using KOH solution (30 wt.%) at 120 °C to suspend the PhC slab. Finally, modified RIE etching was applied without a mask to reduce the thickness of the slab to 265 nm (or 233 nm).

Few-layers h-BN flakes and monolayer $WSe_2$ samples were mechanically exfoliated from bulk crystals onto 290 nm thick $SiO_2/Si$ wafers. $WSe_2$ monolayers were identified under an optical microscope and verified using photoluminescence measurements. Few-layer h-BN flakes were then transferred onto the identified $WSe_2$ monolayer samples using our previously demonstrated capillary-force-assisted clean-stamp transfer method[42]. The h-BN/$WSe_2$ heterostructure was then transferred onto the suspended photonic crystal slab device using the same transfer method.

**Optical measurement**

The Fourier-optics-based spectroscopy has three operating modes: an imaging mode, an optical band-structure (angle-resolved reflection spectroscopy) mode and a spectra analysis mode. For the imaging mode, a broadband emission halogen lamp (QTH10, Thorlabs) with a wavelength range of 400 to 2200 nm was used to provide white light. A CCD camera (16MP high-speed USB 3.0 digital camera, AmScope) was placed at the image-plane after the tube lens to show the image (**Fig. S3**). For the optical band-structure (angle-resolved reflection spectroscopy) mode, the back focal plane of an objective lens (20X Mitutoyo Plan Apo Infinity Corrected Long WD objective, NA = 0.42 or 40X Nikon S Flour high NA objective, NA = 0.9) was projected by a Fourier transform system. A pinhole was placed at the image-plane after the tube lens (f = 200 mm) to regulate the field of view for the light reflected from photonic crystal slab. A slit spectrometer (Princeton Instruments, spectrometer SP300 and silicon camera PIXIS 400) was placed at the K-plane and switched to spectrometer mode for the optical band-structure or imaging mode for K-space imaging. For the spectra analysis mode, the Fourier lens L4 (f = 100 mm) was replaced with an imaging lens L3 (f = 50 mm) to Fourier transform the K-plane to an image-plane. The spectrometer can be used to collect the spectra in the image-plane. (**Fig. S3b**).

Photoluminescence spectroscopy can be achieved by switching flip mirror-1 to the laser mode. The sample was pumped by a wavelength tunable mode-locked Ti:Sapphire laser (Chameleon Ultra II, Coherent) and cascade of OPO (optical parametric oscillator). A high numerical aperture (NA = 0.9) 40X objective lens (40X Nikon S Flour high NA objective) was used to collect the PL signal at room temperature. The PL signal was then sent to the K-plane by the Fourier transform system and imaged at the spectrometer for either K-plane imaging or the PL emission momentum distribution. A short pass filter (FESH0700, Thorlabs) was used to clean the undesired wavelength before the pump laser reached the objective length and a long pass filter (FELH0700, Thorlabs) was used to block the pump power to make the PL signal stand-out.

**Numerical simulation**

Numerical simulations of the far field emission patterns and Purcell enhancement factors are carried out using three-dimensional full-wave finite-difference time-domain methods (Lumerical FDTD Solutions, ANSYS Inc.). The simulation spans in x and y directions encompassing a sample area of 30 by 30 periods. Perfectly matched layers (PMLs) are used as the boundary conditions in all directions. The $X_D$ emission is simulated by placing a z-oriented dipole array on the upper surface of the photonic crystal slab. The radiation phase of each dipole is set randomly. The minimum mesh size is set as 2 nm around the emitting region. To collect the emission profile, a far field pattern monitor is placed 1 μm above the sample surface. To calculate the Purcell enhancement, a control simulation with the same $Si_3N_4$ slab without the holes is carried out. The Purcell enhancement factor is calculated by extracting the emission intensity at 48° for both cases with and without lattices.

Numerical simulations of the angle-resolved reflection spectra mapping are carried out using the finite element method (COMSOL Multiphysics). We selected one unit-cell as the simulation domain and use the periodic (Bloch) boundary condition in x and y directions and the periodic type of port to excite the horizontal magnetic field with various oblique incident angles for the input light. The S-parameter, S11, was used for the reflection spectra.

**Reference**

1. Ye, Z. et al. Probing excitonic dark states in single-layer tungsten disulphide. *Nature* **513**, 214-218 (2014).
2. Zhang, X. X. et al. Magnetic brightening and control of dark excitons in monolayer WSe$_2$. *Nat. Nanotechnol.* **12**, 883-888 (2017).
3. Zhou, Y. et al. Probing dark excitons in atomically thin semiconductors via near-field coupling to surface plasmon polaritons. *Nat. Nanotechnol.* **12**, 856-860 (2017).
4. Park, K. D., Jiang, T., Clark, G., Xu, X. & Raschke, M. B. Radiative control of dark excitons at room temperature by nano-optical antenna-tip Purcell effect. *Nat.Nanotechnol.* **13**, 59-64 (2018).
5. Smoleński, T., Kazimierczuk, T., Goryca, M., Wojnar, P. & Kossacki, P. Mechanism and dynamics of biexciton formation from a long-lived dark exciton in a CdTe quantum dot. *Phys. Rev. B* **91**, 155430 (2015).
6. Heindel, T. et al. Accessing the dark exciton spin in deterministic quantum-dot microlenses. *APL Photonics* **2**, 121303 (2017).
7. Akinwande, D. et al. Graphene and two-dimensional materials for silicon technology. *Nature* **573**, 507-518 (2019).
8. Echeverry, J., Urbaszek, B., Amand, T., Marie, X. & Gerber, I. Splitting between bright and dark excitons in transition metal dichalcogenide monolayers. *Phys. Rev. B* **93**, 121107 (2016).
9. Tang, Y., Mak, K. F. & Shan, J. Long valley lifetime of dark excitons in single-layer WSe$_2$. *Nat. Commun.* **10**, 1-7 (2019).
10. Wang, G. et al. In-plane propagation of light in transition metal sichalcogenide monolayers: optical selection rules. *Phys. Rev. Lett.* **119**, 047401 (2017).
11. Ma, X. et al. Engineering photonic environments for two-dimensional materials. *Nanophotonics* **10**, 1031-1058 (2021).
12. Pyatkov, F. et al. Cavity-enhanced light emission from electrically driven carbon nanotubes. *Nat. Photonics* **10**, 420-427 (2016).
13. Javerzac-Galy, C. et al. Excitonic emission of monolayer semiconductors near-field coupled to high-Q microresonators. *Nano Lett.* **18**, 3138-3146 (2018).
14. Dong, Z. et al. Silicon nanoantenna mix arrays for a trifecta of quantum emitter enhancements. *Nano Lett.* **21**, 4853-4860 (2021).
15. Iyer, P. P. et al. Unidirectional luminescence from InGaN/GaN quantum-well metasurfaces. *Nat. Photonics* **14**, 543-548 (2020).
16. Hsu, C. W. et al. Observation of trapped light within the radiation continuum. *Nature* **499**, 188-191 (2013).
17. Azzam, S. I., Shalaev, V. M., Boltasseva, A. & Kildishev, A. V. Formation of bound states in the continuum in hybrid plasmonic-photonic systems. *Phys. Rev. Lett.* **121**, 253901 (2018).
18. Koshelev, K., Favraud, G., Bogdanov, A., Kivshar, Y. & Fratalocchi, A. Nonradiating photonics with resonant dielectric nanostructures. *Nanophotonics* **8**, 725-745 (2019).
19. Gao, X., Zhen, B., Soljacic, M., Chen, H. & Hsu, C. W. Bound states in the continuum in fiber bragg gratings. *ACS Photonics* **6**, 2996-3002 (2019).

20. Hsu, C. W., Zhen, B., Stone, A. D., Joannopoulos, J. D. & Soljačić, M. Bound states in the vontinuum. *Nat. Rev. Mater.* **1**, 16048 (2016).

21. Bulgakov, E. N. & Maksimov, D. N. Avoided crossings and bound states in the continuum in low-contrast dielectric gratings. *Phys. Rev. A* **98**, 053840 (2018).

22. Friedrich, H. & Wintgen, D. Interfering resonances and bound states in the continuum. *Phys. Rev. A* **32**, 3231 (1985).

23. Lyapina, A., Maksimov, D., Pilipchuk, A. & Sadreev, A. Bound states in the continuum in open acoustic resonators. *J. Fluid Mech.* **780**, 370-387 (2015).

24. Sadreev, A. F., Bulgakov, E. N. & Rotter, I. Bound states in the continuum in open quantum billiards with a variable shape. *Phys. Rev. B* **73**, 235342 (2006).

25. Marinica, D., Borisov, A. & Shabanov, S. Bound states in the continuum in photonics. *Phys. Rev. Lett.* **100**, 183902 (2008).

26. Lee, S. G., Kim, S. H. & Kee, C. S. Bound states in the continuum (BIC) accompanied by avoided crossings in leaky-mode photonic lattices. *Nanophotonics* **9**, 4373-4380 (2020).

27. Sadreev, A. F. Interference traps waves in open system: Bound states in the continuum. *Rep. Prog. Phys.* **84**, 055901 (2021).

28. Guo, Y., Xiao, M. & Fan, S. Topologically protected complete polarization conversion. *Phys. Rev. Lett.* **119**, 167401 (2017).

29. Bogdanov, A. A. et al. Bound states in the continuum and Fano resonances in the strong mode coupling regime. *Adv. Photonics* **1**, 016001 (2019).

30. Rybin, M. V. et al. High-Q supercavity modes in subwavelength dielectric resonators. *Phys. Rev. Lett.* **119**, 243901 (2017).

31. Wang, L. et al. One-dimensional electrical contact to a two-dimensional material. *Science* **342**, 614-617 (2013).

32. Kravtsov, V. et al. Nonlinear polaritons in a monolayer semiconductor coupled to optical bound states in the continuum. *Light Sci. Appl.* **9**, 1-8 (2020).

33. Lee, S. Y., Jeong, T. Y., Jung, S. & Yee, K. J. Refractive index dispersion of hexagonal boron nitride in the visible and near-infrared. *Phys. Status Solidi B* **256**, 1800417 (2019).

34. Johnson, S. G. & Joannopoulos, J. D. Block-iterative frequency-domain methods for Maxwell's equations in a planewave basis. *Opt. Express* **8**, 173-190 (2001).

35. You, Y. et al. Observation of biexcitons in monolayer $WSe_2$. *Nat. Phys.* **11**, 477-481 (2015).

36. Klingshirn, C. F. *Semiconductor optics*. (Springer Science & Business Media, 2012).

37. Huang, L., Xu, L., Rahmani, M., Neshev, D. & Miroshnichenko, A. E. Pushing the limit of high-Q mode of a single dielectric nanocavity. *Adv. Photonics* **3**, 016004 (2021).

38. Yu, H., Cui, X., Xu, X. & Yao, W. Valley excitons in two-dimensional semiconductors. *Natl. Sci. Rev.* **2**, 57-70 (2015).

39. Qiu, L., Chakraborty, C., Dhara, S. & Vamivakas, A. Room-temperature valley coherence in a polaritonic system. *Nat. Commun.* **10**, 1513 (2019).
40. Shvedov, V. G., Hnatovsky, C., Krolikowski, W. & Rode, A. V. Efficient beam converter for the generation of high-power femtosecond vortices. *Opt. Lett.* **35**, 2660-2662 (2010).
41. Tang, Y. et al. Harmonic spin–orbit angular momentum cascade in nonlinear optical crystals. *Nat. Photonics* **14**, 658-662 (2020).
42. Ma, X. et al. Capillary-force-assisted clean-stamp transfer of two-dimensional materials. *Nano Lett.* **17**, 6961-6967 (2017).

**Data availability**

All data that support the findings of this study are available in the main text, figures, and Supplementary Information. They are also available from the corresponding author upon reasonable request.

**Author contributions**

X. M. and S. L. conceived the idea and initiated the project. X.M. performed the numerical simulation with Y.C., Z.J.W., Y.M., and M.C.H. helped. X.M. fabricated devices and performed optical measurements with P.C. and K.K. helped. X.M. analysed the data. K.W. and T.T. grew the bulk boron nitride crystals. X.M. and S.L. wrote the manuscript with X.Q, P.C. and K.K. helped. All authors discussed the results and manuscript. S.L. supervised the project.

**Conflict of Interest**

The authors declare no conflict of interest.

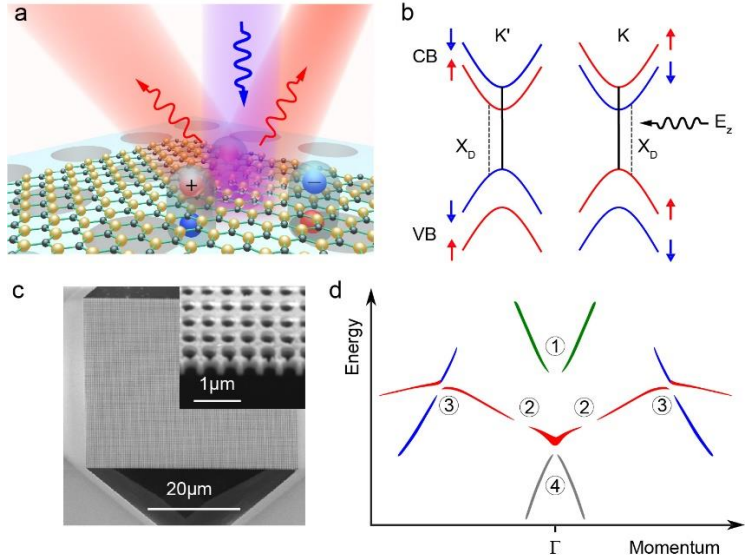

**Figure 1. Dark exciton brightening and directional emission. a**, Schematic of directional emission of dark excitons in the WSe$_2$ monolayer with normally incident pumping light. The dark excitons PL signal can then be directionally emitted through the Friedrich-Wintgen bound states in the continuum (BIC) supported by the same PhC slab. **b**, Split-band configuration of bright and dark exciton states. The optically forbidden transition of dark exciton (X$_D$) is brightened by the converted E$_z$ with enhancement on the top surface of the PhC slab. CB, conduction band; VB, valence band. **c**, A scanning electron microscope (SEM) image of the PhC slab made of silicon nitride (Si$_3$N$_4$). **d**, A sketch of the optical band structure of the PhC slab with three types of the BICs: ① and ④ are the on-Γ symmetry-protected BICs, ② is the off-Γ accidental BIC, and ③ is the Friedrich-Wintgen BIC due to the destructive interference of resonances belonging to different bands (red and blue).

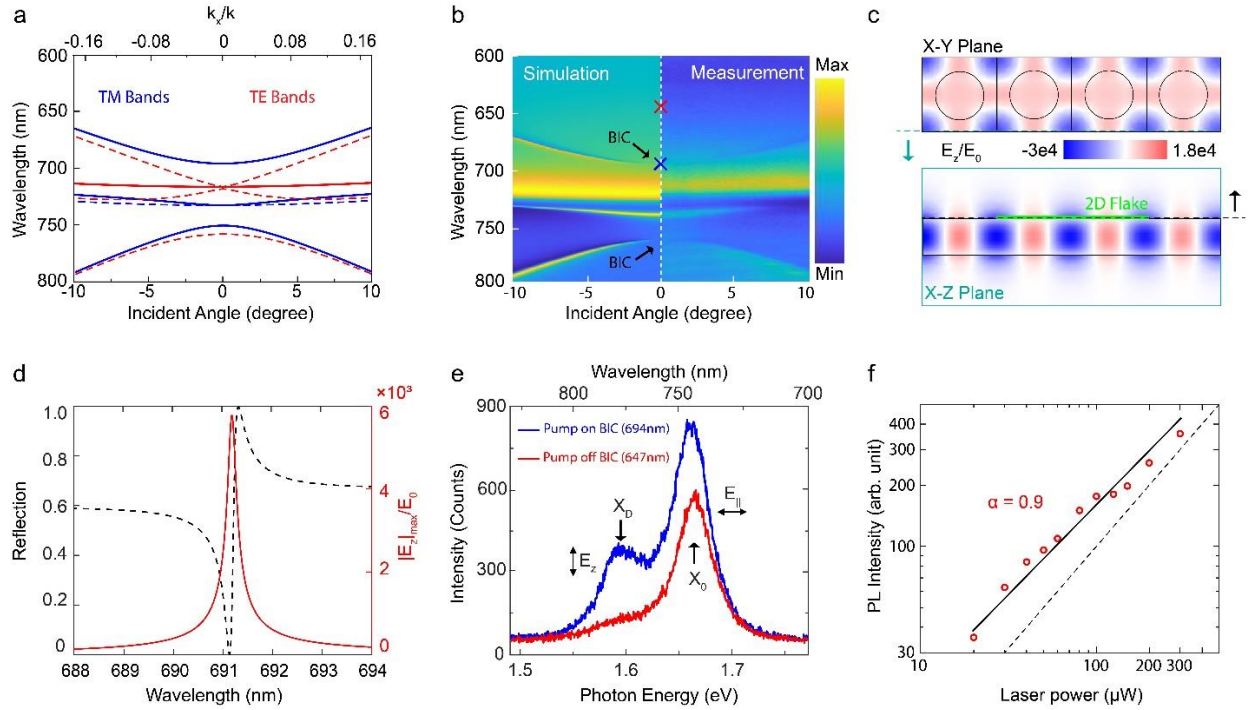

**Figure 2. Brightening of dark excitons with BICs. a**, The band structure of a PhC slab that supports a symmetry-protected on-Γ BIC with a zero degree incident angle. The blue bands are the transverse magnetic mode-like (TM-like) bands where the red bands are the transverse electric mode-like (TE-like) bands. Only four of the bands (solid lines) can be observed under p-polarized incident light. The PhC only supports the on-Γ BIC to illustrate the dark exciton brightening. **b**, The simulated (left) and the measured (right) angle-resolved reflection spectra mapping of the PhC slab. It is clear to see the two on-Γ BICs at wavelengths 694nm and 750nm, respectively. **c**, Electric-field profile $E_z/E_0$ of the on-Γ BICs, plotted on the top surface of the PhC slab (top) and the y = -r/2 slice (bottom). **d**, The reflection spectrum with an oblique incident angle of 3° is shown by the black dashed line, while the maximum local electric field amplitude enhancement ratio on the top of the PhC slab is plotted in red. **e**, The PL spectra of dark excitons and bright excitons. The blue spectrum was taken when the pump laser matched the on-Γ BICs at the wavelength of 694nm (on-BIC) whereas the red spectrum was taken when the pump laser was at the wavelength of 647 nm (off-BIC). **f**, A log-plot of the power dependence of PL intensity of dark excitons. The black line is a fit of the dark exciton emissions exhibiting a linear power dependence. The fitted slop α is 0.9 indicating the PL stem from dark excitons rather than bi-excitons.

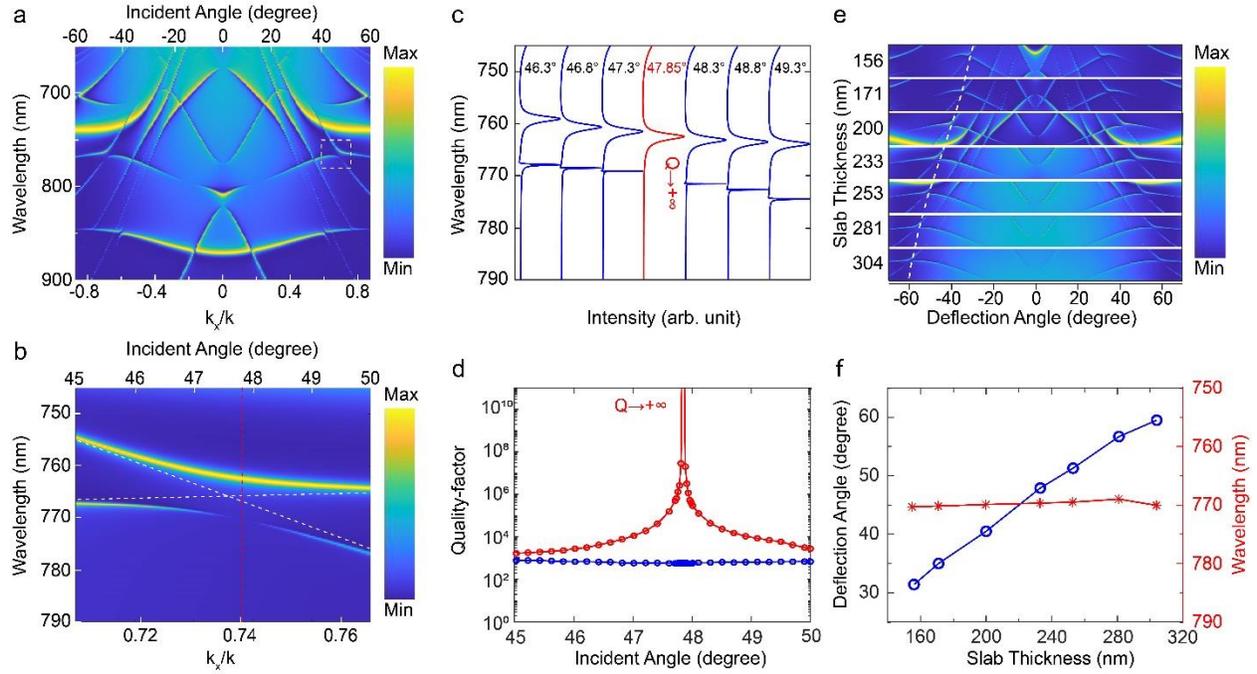

**Figure 3. Tunable Friedrich-Wintgen BICs. a**, Dispersion spectra the PhC that supports BICs. The Friedrich-Wintgen BIC due to the interference of two TM-like bands is highlighted by the white dashed box. **b**, A close look of the Friedrich-Wintgen BIC at a wavelength of 770 nm and oblique incident angle of 47.85° (red line). The white dashed line indicates the original of the two modes. **c**, Spectra at a series of different incident angles. Avoided crossing and linewidth vanishing of the lower branch band at 47.85° are observed due to the interference between the two modes. **d**, Quality factors (Q-factors) of two bands that form the Friedrich-Wintgen BIC as a function of the oblique incident angle. The blue circles and the red circles represent the Q-factors of the upper branch and the lower branch of the avoided crossing bands, respectively. Q-factors of the lower branch are rapidly increasing when the oblique incident angle approaches 47.85° where the Friedrich-Wintgen BIC occurred. **e**, Friedrich-Wintgen BIC modes for $X_D$ directional emission are tunable for different deflection angles. The thicker the PhC slab, the higher the deflection angle of the Friedrich-Wintgen BIC emission channel. **f**, Quasi-linear relations (blue) between the deflection angle by Friedrich-Wintgen BICs and the PhC slab thickness. The Friedrich-Wintgen BIC can be tuned in the momentum space from 31.4° to 59.5° and maintain the wavelength close to ~770 nm (red).

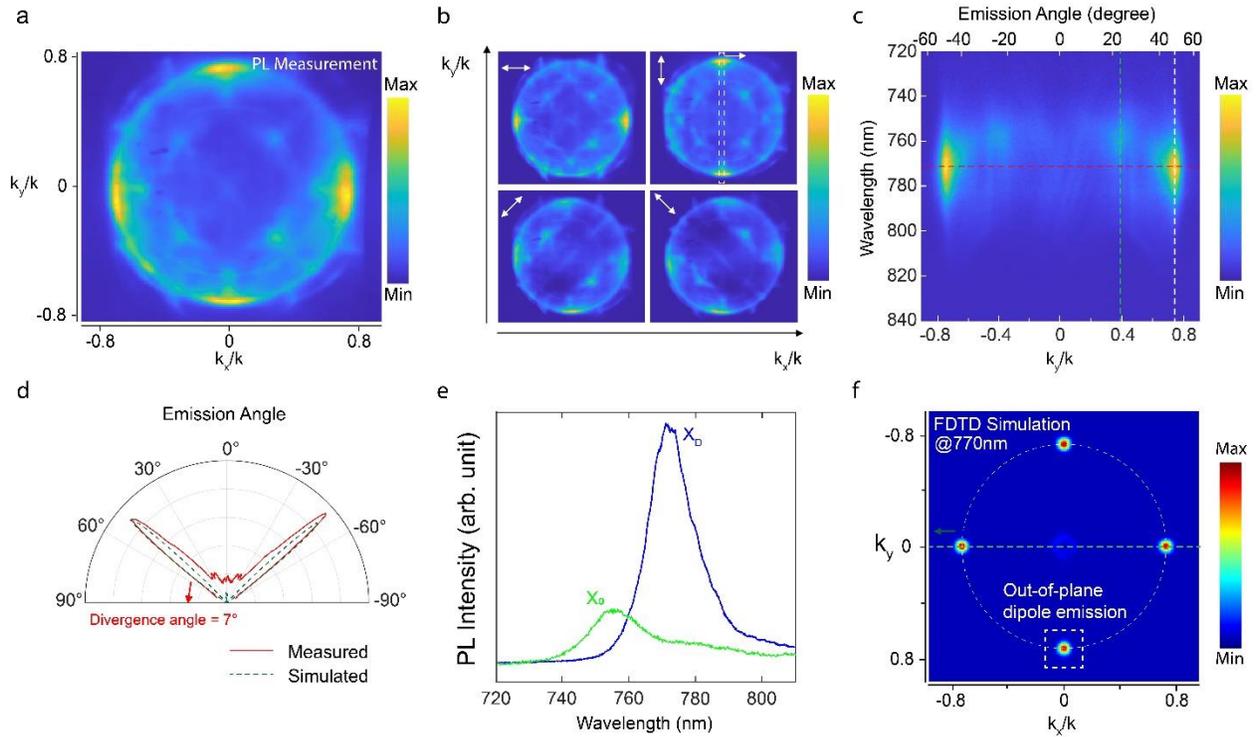

**Figure 4. Directionality control of dark excitons. a**, PL emission momentum distribution mapping with $k_x$- and $k_y$ axis. Four shining emission spots, located at the Γ-X line and $k_x$ or $k_y = 0.74$ with $C_4$ symmetry, are clearly visible. **b**, Polarization analysis of the PL emission momentum distribution in (**a**). The four shining emission spots show radial polarization indicating they are from dark excitons. **c**, Angle-resolved PL emission spectra mapping extracted from y-polarized PL emission momentum distribution mapping in (**b**). It is clear to see the dark exciton directional emissions have small divergence angles at wavelengths of around 772 nm and towards oblique emission angles of around 48°. **d**, The measured (red solid line) and FDTD simulated (dark green dashed line) PL intensity as a function of the in-plane momentum ($k_y/k$) along the y-direction. The full-width-half-maximum (FWHM) of the measured $X_D$ emission lobes is 7° indicating the ultra-low divergence angle of the directional emission. **e**, Spectra extracted from oblique angles of 48° and 23° for the dark exciton emission and the bright exciton emission, respectively. **f**, The simulated PL emission momentum distribution by the Lumerical FDTD. Four shining emission spots show high correspondence to the measured PL emission pattern in (**a**).

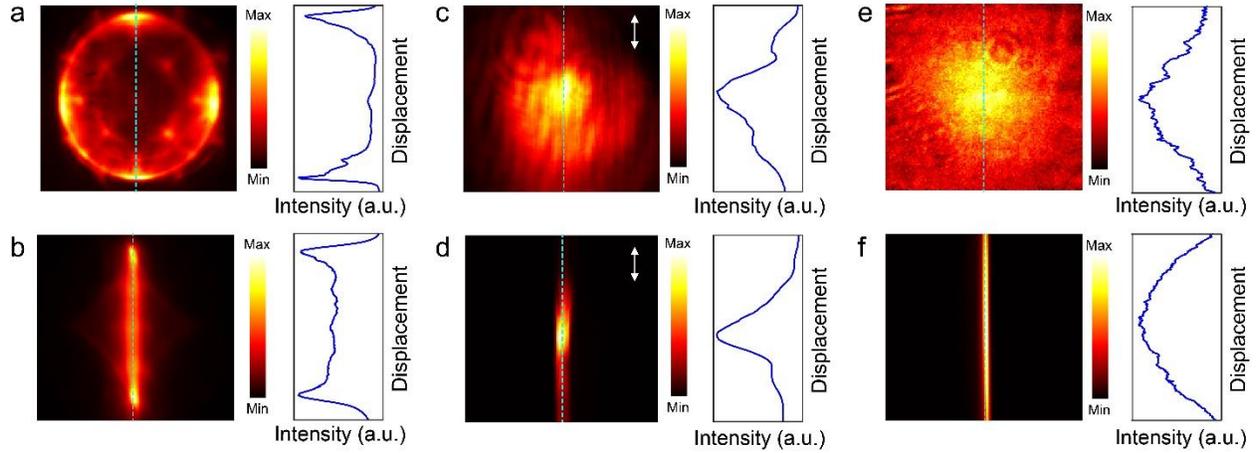

**Figure 5. Room-temperature coherence of the directional emission. a-b**, Momentum distribution and spatial interference enabled by a cylindrical lens for the directional emission of dark excitons supports an interference pattern **(b)**. The destructive interference in the middle of **(b)** is due to a π-phase shift between the left and right parts of the light field in **(a)**, showing that the direction emission is strongly coherent at room temperature. **c-d**, Momentum distribution and spatial interference for coherent laser beam. The center of **(d)** is brighter than that of **(c)**, which is caused by constructive interference. **e-f**, Momentum distribution and spatial interference for incoherent bright exciton emission of monolayer $WSe_2$ on $SiO_2$/Si substrate. The intensity profile of **(f)** is similar to that of **(e)**, because the bright exciton lots its valley coherence at room temperature.